\documentclass[reprint, superscriptaddress, secnumarabic, amssymb, nobibnotes, aps, prl]{revtex4-1}

\usepackage{graphicx}
\usepackage{epstopdf}
\usepackage[T1]{fontenc}
\usepackage[latin9]{inputenc}
\usepackage{amsbsy}
\usepackage{gensymb}
\usepackage[T1]{fontenc}
\usepackage[latin9]{inputenc}
\usepackage{amsmath}
\usepackage{amssymb}
\usepackage{bbm}
\usepackage{braket}
\usepackage{xcolor}
\allowdisplaybreaks
\usepackage{graphicx}
\usepackage[colorlinks=true]{hyperref}  
\hypersetup{
    bookmarks=true,         
    unicode=false,          
    pdftoolbar=true,        
    pdfmenubar=true,        
    pdffitwindow=false,     
    pdfstartview={FitH},    
    pdftitle={Muon spin spectroscopy study of the noncentrosymmetric superconductor TaOs},    
    pdfauthor={D. Singh, Sajilesh K.P, A. D. Hillier, R. P. Singh},     
    pdfsubject={},   
    pdfcreator={},   
    pdfproducer={}, 
    pdfkeywords={} {} {}, 
    pdfnewwindow=true,      
    colorlinks=true,       
    linkcolor=blue, 
    citecolor=blue,        
    filecolor=magenta,      
    urlcolor=blue           
} 
\usepackage[normalem]{ulem}

\newcommand{\equref}[1]{Eq.~(\ref{#1})}

\newcommand{\figref}[1]{Fig.~\ref{#1}}

\newcommand{\tableref}[1]{Table~\ref{#1}}

\renewcommand{\approx}{\simeq}

\begin{document}
\title{\textrm{Superconducting and normal state properties of noncentrosymmetric superconductor NbOs$_{2}$ investigated by muon spin relaxation and rotation}}
\author{D. Singh}
\affiliation{Indian Institute of Science Education and Research Bhopal, Bhopal, 462066, India}
\author{Sajilesh K.P}
\affiliation{Indian Institute of Science Education and Research Bhopal, Bhopal, 462066, India}
\author{Sourav Marik}
\affiliation{Indian Institute of Science Education and Research Bhopal, Bhopal, 462066, India}
\author{A. D. Hillier}
\affiliation{ISIS facility, STFC Rutherford Appleton Laboratory, Harwell Science and Innovation Campus, Oxfordshire, OX11 0QX, UK}
\author{R. P. Singh}
\email[]{rpsingh@iiserb.ac.in}
\affiliation{Indian Institute of Science Education and Research Bhopal, Bhopal, 462066, India}

\date{\today}
\begin{abstract}
\begin{flushleft}

\end{flushleft}
Noncentrosymmetric superconductors with $\alpha$-manganese structure has attracted much attention recently, after the discovery of time-reversal symmetry breaking in all the members of Re$_{6}$X (X = Ti, Hf, Zr) family. Similar to Re$_{6}$X, NbOs$_{2}$ also adopts $\alpha$-$Mn$ structure and found to be superconducting with critical temperature $T_{c}$ $\approx$ 2.7 K. The results of the resistivity, magnetization, specific heat and muon-spin relaxation/rotation measurements show that NbOs$_{2}$ is a weakly coupled type-II superconductor. Interestingly, the zero-field muon experiments indicate that the time-reversal symmetry is preserved in the superconducting state. The low-temperature transverse-field muon measurements and the specific heat data evidence an conventional isotropic fully gapped superconductivity. However, the calculated electronic properties in this material show that the NbOs$_{2}$ is positioned close to the band of unconventionality of the Uemura plot, indicating that NbOs$_{2}$ potentially borders an unconventional superconducting ground state.    
\end{abstract}
\maketitle
\section{Introduction}
The discovery of exotic superconducting properties in the heavy fermion noncentrosymmetric superconductor (NCS) CePt$_{3}$Si \cite{EB}, sparked renewed research interest both from both experimental and theoretical perspectives, to understand the role of structural asymmetry in superconductivity. Theoretically, in the Bardeen-Cooper-Schrieffer (BCS) superconductors, the pairing state is protected by the crystal inversion symmetry. When a superconductor has an inversion center, the pairing states can be unambiguously classified into degenerated states of spin-singlet (even parity) and spin-triplet (odd parity). However, this scenario changes in noncentrosymmetric superconductors, where the lack of inversion symmetry induces odd parity antisymmetric spin-orbital coupling (ASOC). A strong ASOC results in mixed parity states and develops complicated spin structures \cite{EB1, LP, PAF}. Such an admixture of spin states reflects in a variety of unique properties, such as: anomalous upper critical field values \cite{EB2,NK,IS}, line nodes \cite{IB,EB3}, and magnetoelectric effects \cite{VM,SF} etc. Besides, recently it was proposed that some NCS can also be a potential candidate for topological superconductivity, of which known examples are PbTaSe$_{2}$ \cite{GB1,SYG} and BiPd \cite{ZSM,MN}, which shows topological surface states and potentially lead to signatures of Majorana fermions.\\
Since CePt$_{3}$Si, several more heavy fermion superconductors have been found to exhibit a wide variety of unusual superconducting properties \cite{EB1}. But it was soon realized that heavy fermion noncentrosymmetric superconductors are not the simplest of systems to study. In these superconductors, additional complications come due to the presence of strong electronic correlation effects and quantum criticality, which often hinders the research aimed to understand the interplay between crystal symmetry and superconductivity. As a result, many new noncentrosymmetric superconductors with weak correlation have been targeted to look for singlet-triplet mixing. Some of these compounds are LaNiC$_{2}$ \cite{ADH,ADH1}, Li$_{2}$Pd$_{3}$B \cite{KTP,PBT,PBT1,MNY,HTK,HQY}, Li$_{2}$Pt$_{3}$B \cite{PBT1,MNY,HTK,HQY,MNY1,HTM1}, Ru$_{7}$B$_{3}$ \cite{LFH}, Nb$_{0.18}$Re$_{0.82}$ \cite{ABK}, Re$_{3}$W \cite{PKB}, Y$_{2}$C$_{3}$ \cite{JCM}, Mo$_{3}$Al$_{2}$C \cite{MLC,MLC1}. A particularly interesting example is Li$_{2}M_{3}$B (M = Pd, Pt) \cite{HTK,HQY,MNY1,HTM1,TMH,RKI,MMD,HQY1,STT,ASY,DCP,GED}. Li$_{2}$Pd$_{3}$B exhibits a isotropic fully gapped superconductivity \cite{MNY,HTK, HQY}, whereas the gap of Li$_{2}$Pt$_{3}$B has line nodes \cite{HQY,MNY1,HTM1}. Here the substitution of Pd with Pt element strengthens the ASOC and enhances the relative pairing mixing ratio with a dominant spin-triplet component \cite{HQY}. Hence, the change from fully gapped to nodal superconductivity is directly dependent on the magnitude of ASOC. Indeed, $^{11}B$ NMR experiments also conclude that the Cooper pair included about 60$\%$ of the spin-triplet pairing in Li$_{2}$Pt$_{3}$B.
Accordingly, many noncentrosymmetric superconductors with wide variety of SOC have been studied, yet most of them such as LaMSi$_{3}$ (M = Pd, Pt)\cite{MSA3}, LaMSi$_{3}$ (M = Rh, Ir)\cite{VKA1,VKA2}, BiPd \cite{KMS,XBY}, Nb$_{0.18}$Re$_{0.82}$ \cite{ABK}, Re$_{3}$W \cite{PKB} found to be compatible with fully gapped superconductivity. Thus, the importance of ASOC strength in the degree of mixing ratio is not fully understood, and therefore requires to find more examples where the effects of modified spin-orbit coupling on the singlet-triplet mixing can be studied directly.\\ 
Another noteworthy feature of unconventional superconductivity is time-reversal symmetry (TRS) breaking, which is an extremely rare phenomenon has been found only in few noncentrosymmetric superconductors, such as: LaNiC$_{2}$ \cite{ADH}, La$_{7}$Ir$_{3}$ \cite{JAT}, SrPtAs \cite{PKB1} and Re$_{6}$X (X = Ti,Hf,Zr) \cite{RPS,DSJ,DSJ1}. It is the latter family of compounds that are particularly interesting owing to the frequent occurrence of TRS breaking among all the members. In Re$_{6}$X, the TRS breaking effects seem very similar irrespective of the substitution at X-sites, suggesting a negligible effect on the ASOC and hints towards its common origin. Interestingly, the band-structure calculations of the Re-based binary alloys \cite{r1,r2}, indicate that the density of states (DOS) at the Fermi level to be dominated by the d-bands of Re and could be the critical factor leading to TRS breaking. Indeed, recent observation of TRS breaking in pure Re metal (centrosymmetric), strongly suggests that the sizable Re spin-orbit coupling to be the origin of TRS breaking in the Re$_{6}$X family \cite{TMC}.
However, to strengthen the above conclusion one need to study other Re-free materials, whilst still having the same $\alpha$-$Mn$ type structure. In this context, we report a detailed investigation of the superconductivity of NCS NbOs$_{2}$, isostructural to the Re$_{6}$X family. NbOs$_{2}$ exhibits bulk superconductivity around 2.7 K \cite{BT,BT1}, characterized via electrical resistivity, specific heat, magnetic susceptibility, and muon spectroscopy. The principal goal of the present work is to study the superconducting properties of NbOs$_{2}$ and search for TRS breaking in a Re-free material with $\alpha$-$Mn$ structure.
\section{Experimental Details}
The polycrystalline samples of NbOs$_{2}$ were prepared by arc melting a stoichiometric mixture of Nb (99.95 $\%$) and Os (99.95 $\%$) under high purity argon gas atmosphere on a water cooled copper hearth. The as-cast ingot was flipped and remelted several times to ensure the phase homogeneity. The observed mass loss was negligible. The sample was then sealed inside an evacuated quartz tube and annealed at 800 $\degree$C for one week. To verify the phase purity we  performed room temperature (RT) powder x-ray diffraction (XRD) using a X'pert PANalytical diffractometer (Cu-K$_{\alpha 1}$ radiation, $\lambda$ = 1.540598 \text{\AA}).
The superconducting properties of NbOs$_{2}$ were measured using magnetization $M$, ac susceptibility $\chi_{ac}$, electrical resistivity $\rho$, specific heat $C$,  and muon relaxation/rotation ($\mu$SR) measurements. The dc magnetization and ac susceptibility data were collected using a Quantum Design superconducting quantum interference device (SQUID). The electrical resistivity and specific heat measurements were performed on a Quantum Design physical property measurement system (PPMS). The $\mu$SR measurements were conducted at the ISIS Neutron and Muon facility, in STFC Rutherford Appleton Laboratory, United Kingdom using the MUSR spectrometer. The powdered NbOs$_{2}$ sample was mounted on a silver holder and placed in a sorption cryostat, which we operated in the temperature range 0.3 K - 3.0 K. Zero-field muon spin relaxation (ZF-$\mu$SR) and the transverse-field muon spin rotation (TF-$\mu$SR) measurements were performed on the MuSR beam line at the ISIS pulsed muon source. A full description of the $\mu$SR technique may be found in Ref. \cite{SLL}. In ZF-$\mu$SR, the contribution from the stray fields at the sample position due to neighbouring instruments and the Earth's magnetic field is cancelled to within $\sim$ 1.0 $\mu$T using three sets of orthogonal coils. TF-$\mu$SR measurements are performed to measure the temperature dependence of the  magnetic-penetration depth $\lambda$(T). $\lambda^{-2}$(T) is proportional to the superfluid density, and can provide information on the symmetry of the gap structure.

\section{Results and Discussion}
\subsection{Normal and superconducting state properties} 
\begin{figure}
\includegraphics[width=1.0\columnwidth]{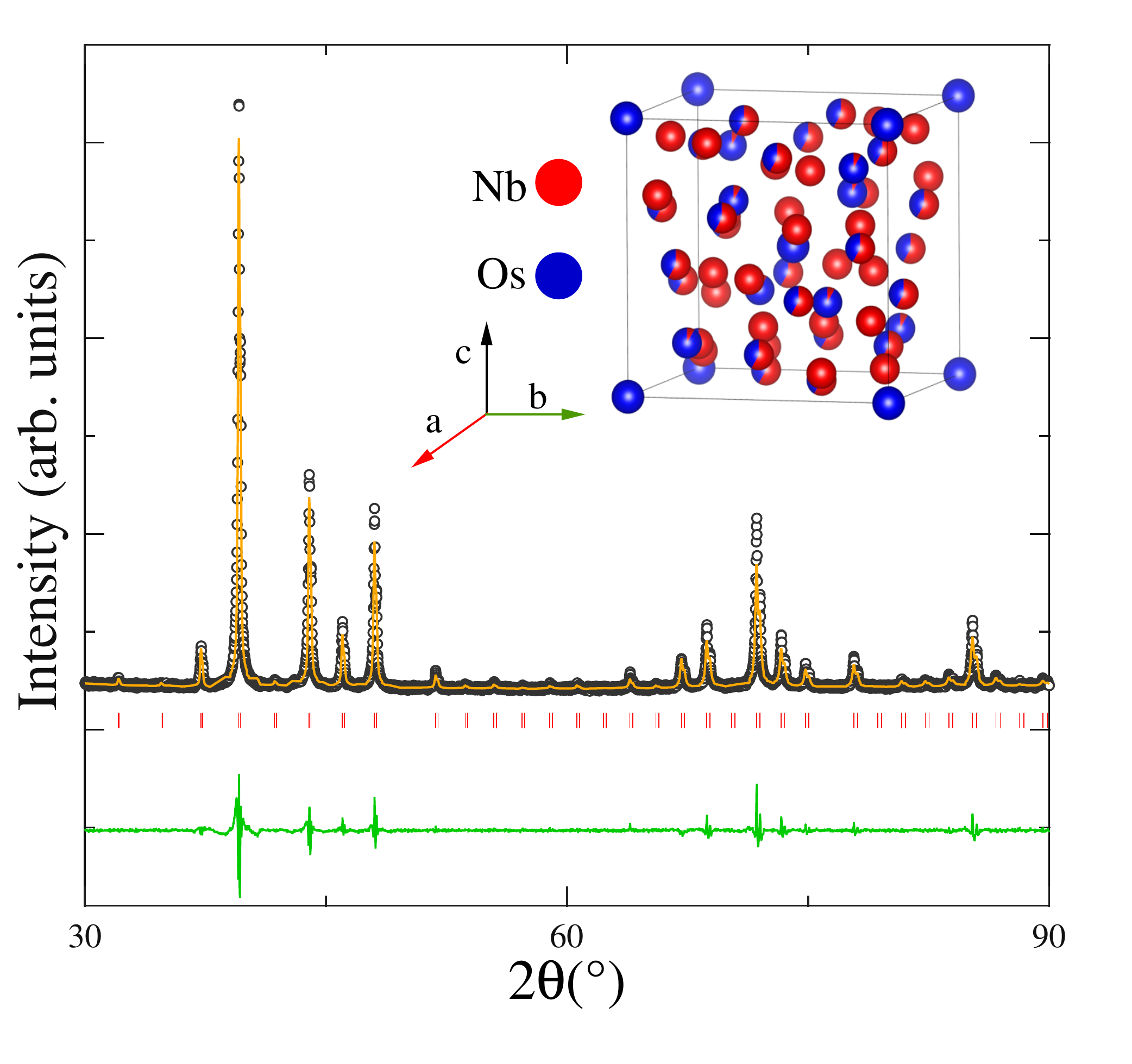}
\caption{\label{Fig1:xrd}  The powder x-ray diffraction pattern of NbOs$_{2}$ at room temperature. The line is a Rietveld refinement to the data.  The inset shows the crystal structure of NbOs$_{2}$.}
\end{figure}

Figure \ref{Fig1:xrd} shows the room temperature powder X-ray diffraction pattern. Rietveld refinement of the RT XRD pattern shows that our sample is in the single phase, crystallizing with a cubic $\alpha$-$Mn$ structure (space group $I \bar{4}3m$) with lattice cell parameter a = 9.654(3) \text{\AA}. A schematic view of the crystal structure of NbOs$_{2}$ is shown in the inset of \figref{Fig1:xrd}.  Refined lattice parameters are shown in \tableref{Crystal structure} and are in good agreement with the published literature \cite{BT}. In the crystal structure except for the Nb(1) site (2a), no other sites have an inversion center. The sample used here is found to have a refined stoichiometry of NbOs$_{1.98}$(3), close to the target composition NbOs$_{2}$, with the possibility of some site mixing between Nb and Os sites \cite{CT}.

\begin{figure}[t]
\includegraphics[width=1.0\columnwidth]{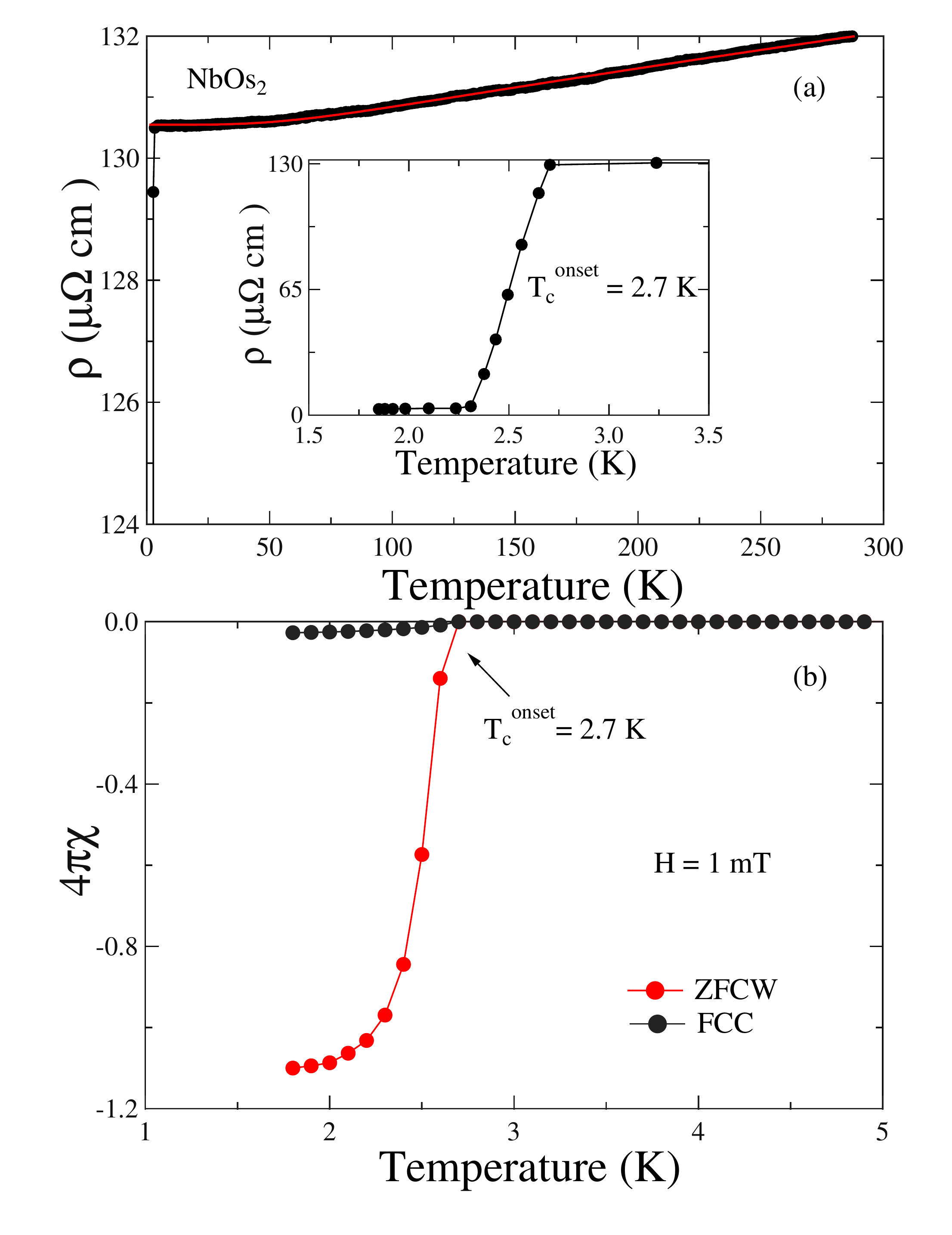}
\caption{\label{Fig2:ZFC} (a) Temperature dependence of resistivity over the range 1.8 K $ \leq $ T $ \leq $ 300 K, with the inset showing a drop in resistivity at the superconducting transition, T$ _{c}^{onset} $ = 2.7 $\pm$ 0.2 K. The normal state resistivity fitted with the Bloch-Gr$\ddot{u}$neisen model shown by solid red line.(b) Temperature dependence of the $\chi$(T) is shown over the range of 1.8 K $ \leq $ T $ \leq $ 5 K, displaying strong diamagnetic signal around $T_{c}^{onset}$ = 2.7 $\pm$ 0.1 K in H = 1 mT.}
\end{figure}

\begin{table}[t]
\caption{Crystal structure parameters obtained from the Rietveld refinement of the room temperature powder x-ray diffraction of NbOs$_{2}$}
~~~Structure~~~~~~~Cubic\\
 ~~~Space group~~~~$I \bar{4}3m$\\
Lattice parameters\\
~~~~~~\textit{a}(\text{\AA})~~~~~~~~~~~~~9.654(3)\\
\begin{center}
\label{Crystal structure}
\begin{tabular*}{1.0\columnwidth}{l@{\hspace{-7ex}}cccc}\hline\hline
Atomic Coordinates\\
\hline
\\[0.5ex]
Atom &Wyckoff position & x & y & z\\                                  
Nb1  &~ 2a & 0 & 0 & 0\\
Os1  &~ 8c & 0.322(1) & 0.322(1) &0.322(1)\\           
Nb2  &~ 8c & 0.322(1) & 0.322(1) & 0.322(1)\\
Os2  &~ 24g & 0.3536(6) & 0.3536(6) & 0.0371(8)\\
Nb3  &~ 24g & 0.3536(6) & 0.3536(6) & 0.0371(8)\\ 
Os3  &~ 24g & 0.0912(3) & 0.0912(3) & 0.2828(4)\\                       
\\[0.5ex]
\hline\hline
\end{tabular*}
\par\medskip\footnotesize
\end{center}
\end{table} 
Temperature dependence of the electrical resistivity $\rho(T)$ in the temperature range 1.8 K to 300 K in zero applied magnetic field is shown in \figref{Fig2:ZFC} (a). The residual resistivity ratio (RRR) is 1.02(1). This low value for the RRR is comparable to other $\alpha$-$Mn$ structure compounds such as Nb$_{0.18}$Re$_{0.82}$  ($\approx$1.3)\cite{ABK}, Re$_{3}$W ($\approx$1.15) \cite{PKB}, Re$_{6}$Hf ($\approx$1.08) \cite{DSJ2}, Re$_{6}$Zr ($\approx$1.09) \cite{DAM}, and Re$_{3}$Ta ($\approx$1.04) \cite{JAT1}. These compounds are widely known to have strong electronic scattering, with a large residual resistivity due to occupational site disorder. This means that the resistivity of the materials with $\alpha$-$Mn$ structure is sensitive to disorder which is responsible for the poor conductivity. A superconducting transition can be seen clearly in the inset of  \figref{Fig2:ZFC}(a) with a onset temperature $T_{c}^{onset}$ = 2.7 $\pm$ 0.2 K.\\
The normal-state resistivity for a non-magnetic metallic crystalline solid is analyzed using the Bloch-Gr$\ddot{\mathrm{u}}$neisen (BG) model, which describes the resistivity arising due to electrons scattering from acoustic phonons. The temperature dependence of the resistivity,  $\rho$(T), is modeled as 
\begin{equation}
 \rho(T) = \rho_{0} + \rho_{BG}(T)
\label{para2}
\end{equation}
where $\rho_{0}$ is the residual resistivity due to the defect scattering and is essentially temperature independent whereas $\rho_{BG} $ is the BG resistivity given by \cite{GG}
\begin{equation}
 \rho_{BG}(T) = C\left(\frac{T}{\Theta_{R}}\right)^{n}\int_{0}^{\Theta_{R}/T}\frac{x^{n}}{(e^{x}-1)(1-e^{-x})}dx . 
\label{para3}
\end{equation}
Here $\Theta_{R}$ is the Debye temperature obtained from resistivity measurements, while C is a material dependent pre-factor and n takes values between 2-5 depending on the nature of the electron scattering \cite{ABA}. A fit to the data employing this model is shown in \figref{Fig2:ZFC}(a) and gave n = (4.2 $ \pm $ 0.2),  Debye temperature $ \Theta_{R} $ =  (376 $ \pm $ 2) K, C = (6.2 $ \pm $ 3.2) $\mu\Omega $cm and residual resistivity $\rho_{0} $ = (130 $ \pm $ 2) $ \mu\Omega $cm. The value of the Debye temperature $\Theta_{R}$ is close to that extracted from the specific heat data.\\
Figure \ref{Fig2:ZFC}(b) displays the temperature dependence of magnetic susceptibility $\chi(T)$ measured in an applied field of 1 mT. Both zero-field cooled warming (ZFCW) and field-cooled cooling (FCC) regimes show a sharp diamagnetic transition around $T_{c}^{onset}$= 2.7 $\pm$ 0.1 K, indicating the onset of bulk superconductivity. In the FCC data, the superconductor does not return to a full Meissner state, indicating strong flux pinning. The calculated Meissner volume fraction (4$\pi$$\chi$) is close to 100 $\%$, suggesting complete superconductivity within the sample.\\
\begin{figure*}[t]
\includegraphics[width=2.0\columnwidth]{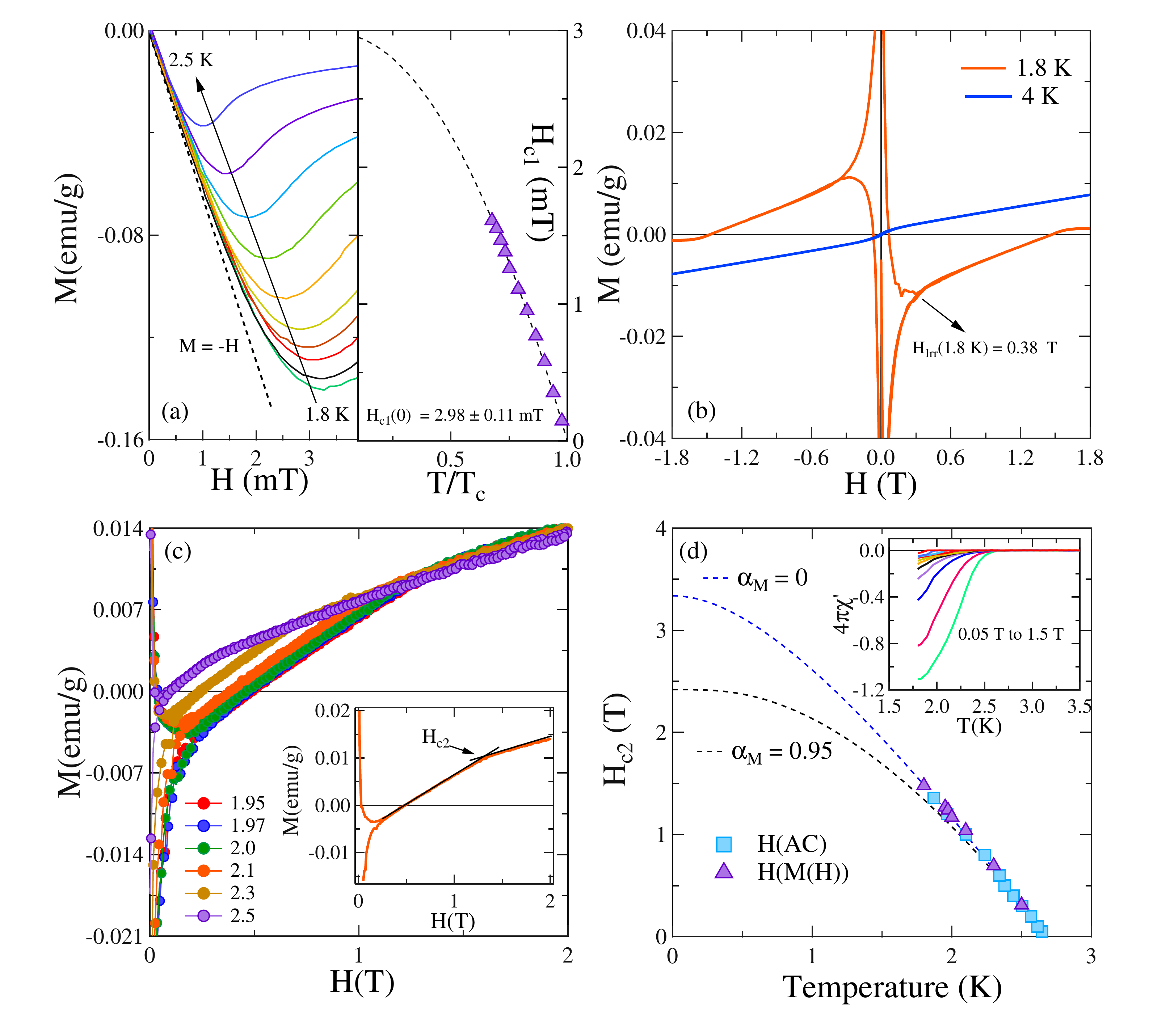}
\caption{\label{Fig3:hc1} (a) The M(H) curves at different fixed temperatures as a function of applied magnetic fields. The lower critical field $H_{c1}$ is determined by the GL formula is 2.98 $\pm$ 0.11 mT. (c) Zero-field cooled magnetic isotherms at T = 1.8 K and T = 4.0 K. (c) Magnetization versus magnetic field is shown at several fixed temperatures. The inset shows the detailed view of M(H) curve, where $H_{c2}$ is determined from the discontinuity in the gradient. (d) Upper critical field versus temperature of NbOs$_{2}$ determined from the high-field M(H) curves and ac susceptibility data. The black and blue dotted line shows the prediction of $H_{c2}$(0) using Eq. \eqref{eqn1:hc2}. The inset shows the ac susceptibility data measured at different applied magnetic field.}
\end{figure*}
To estimate the lower critical field $H_{c1}$, low-field magnetization data $M(H)$ curves were measured at different fixed temperatures from 1.8 K to 2.5 K as shown in \figref{Fig3:hc1}(a). The $M(H)$ data follows a linear relation with the applied magnetic field below $H_{c1}$, whereas, above this point, magnetization starts to deviate from the straight line behavior due to flux penetration. The point of deviation was computed for different isotherm curves to obtain the temperature dependence of $H_{c1}$(T), as presented in the \figref{Fig3:hc1}(a). The resulting graph is modeled using the Ginzburg-Landau relation
\begin{equation} 
 H_{c1}(T) = H_{c1}(0)\left[1-\left(\frac{T}{T_{c}}\right)^{2}\right].
\label{eqn1:hc1}
\end{equation} 
When fitted to the experimental data, it yields $H_{c1}$(0) = 2.98 $\pm$ 0.11 mT.\\
Figure \ref{Fig3:hc1}(b) presents the high-field magnetization hysteresis loop collected in the superconducting state at T = 1.8 K, in the magnetic field range below $\pm$ 2 T. The magnetic behavior clearly corresponds to a type-II superconductor. The irreversibility field, $H_{\mathrm{Irr}}$, is defined as the field where the magnetic hysteresis collapses.  At a given temperature for field $H < H_{\mathrm{Irr}}$, the magnetization is irreversible owing to pinned vortices, whereas for $H > H_{\mathrm{Irr}}$, it becomes reversible as the applied field is strong enough to depin the vortices. As shown in \figref{Fig3:hc1}(b) at T = 1.8 K, $H_{\mathrm{Irr}}$ = 0.38 $\pm$ 0.02 T. The magnetization data were also collected in the normal state at T = 4 K and we conclude that there is no evidence for a magnetic impurity phase in our sample.\\
The magnetic hysteresis loop at different temperatures is displayed in \figref{Fig3:hc1}(c). The hysteresis in magnetization $\Delta$M decreases with increasing temperature and magnetic field, characteristic of a conventional type-II superconductor. The gradient has a clear discontinuity at a field that is identified as the upper critical field $H_{c2}$ [see inset \figref{Fig3:hc1}(c)]. The resulting values of $H_{c2}$ are determined using this discontinuity at different magnetic isotherms  plotted in \figref{Fig3:hc1}(d) (solid triangle markers). The temperature dependence of $H_{c2}$(T) is also determined from the ac susceptibility measurements (solid square markers \figref{Fig3:hc1}(d)), where the shift in $T_{c}$ is evaluated under different applied magnetic fields up to 1.5 T [see inset of \figref{Fig3:hc1}(d)]. The $H_{c2}$ vs. $T$ curve can be described by the Werthamer-Helfand-Hohenberg (WHH) model \cite{EH,NRW} by including the orbital breaking, the effect of Pauli spin paramagnetism ($\alpha$) and spin-orbit scattering parameter ($\lambda_{SO}$). The value of  $\alpha$ measures the relative strengths of the orbital and Pauli-limiting field, while $\lambda_{SO}$ is dominated by the atomic numbers of the elements. In the WHH model
\begin{multline}
\mathrm{ln}\left(\frac{1}{t}\right)= \left(\frac{1}{2}+\frac{i\lambda_{SO}}{4\gamma}\right)\psi\left(\frac{1}{2}+\frac{\bar{h}+\frac{\lambda_{SO}}{2}+{i\gamma}}{2t}\right)\\ +\left(\frac{1}{2}-\frac{i\lambda_{SO}}{4\gamma}\right)
 \psi\left(\frac{1}{2}+\frac{\bar{h}+\frac{\lambda_{SO}}{2}+{i\gamma}}{2t}\right)-\psi\left(\frac{1}{2}\right) ,
\label{eqn1:hc2}
\end{multline}
where t  is the reduced temperature $T/T_{c}$, $\psi$ is the diagamma function,  $\gamma^{2} = (\alpha\bar{h})^{2}-(\frac{\lambda_{SO}}{2})^{2}$, and $\bar{h}$ is the dimensionless form of  the upper critical field given by $\bar{h}$ = (4$\pi^{2})(H_{c2}|dH_{c2}(T)/dT|_{T=T_{c}})$.\\ 
The Maki parameter which measures the relative strengths of the orbital and Pauli-limiting field is calculated using the relation
\begin{equation}
\alpha_{M} = \sqrt{2}\frac{H_{C2}^{orb}(0)}{H_{C2}^{p}(0)}.
\label{eqn1:maki}
\end{equation}
For a type-II superconductor in the dirty limit, the orbital limit of the upper critical field $H_{c2}^{orbital}$(0) is given by the WHH expression  by  
\begin{equation}
H_{c2}^{orbital}(0) = -0.693T_{c}\left.\frac{-dH_{c2}(T)}{dT}\right|_{T=T_{c}}
\label{eqn1:orbit}
\end{equation}
For initial slope of  $-1.80 \pm 0.04 $ T K$^{-1}$ near $T_{c}$ calculated from $H_{c2}$-T plot, it gives $H_{c2}^{orbital}$(0) = (3.36 $\pm$ 0.07) T. The Pauli limiting field is given by $H_{c2}^{p}$(0) = 1.86$T_{c}$ = (5.02 $\pm$ 0.18) T.  The Maki parameter is then calculated  to be $\alpha_{M}$ = 0.95.\\
Figure \ref{Fig3:hc1}(d) show the temperature dependence of $H_{c2}$ for two cases:  $\alpha_{M}$ = 0.95, $\lambda$ = 0 and $\alpha_{M}$ = 0, $\lambda$ = 0, displayed by dotted black and blue lines. It is clear from the graph that the measured data is best described with $\alpha_{M}$ = 0, whereas the calculation with $\alpha_{M}$ = 0.95 fails to account for the experimental data. This certainly implies that $H_{c2}$ is limited by orbital critical field and the Pauli limiting appears to have a limited effect if any at all.\\
The value for $H_{c2}$ can therefore be calculated using the relation \cite{MAKI}
\begin{equation}
H_{c2} = \frac{H_{c2}^{orbital}}{\sqrt{1+\alpha_{M}^{2}}},
\label{eqn1:hc3}
\end{equation}
where it is considered $\lambda$ = 0. For $\alpha_{M}$ = 0, $H_{c2}$ = $H_{c2}^{orbital}$(0) = (3.36 $\pm$ 0.07) T.\\
The value for $H_{c2}$ used to determine the Ginzburg Landau coherence length $\xi_{GL}$ using the relation \cite{tin}
\begin{equation}
H_{c2}(0) = \frac{\Phi_{0}}{2\pi\xi_{GL}^{2}} ,
\label{eqn2:up}
\end{equation}   
where $\Phi_{0}$ (= 2.07 $\times$ 10$^{-15}$ T m$^{2}$) is the magnetic flux quantum. For $H_{c2}$(0) = (3.36 $\pm$ 0.07) T, we find $\xi_{GL}(0)$ = (99 $\pm$ 1)  \text{\AA}.\\
Consequently, the values of $H_{c1}$ and $\xi_{GL}$ can be employed to calculate the Ginzburg-Landau penetration depth $\lambda_{GL}$(0) from the expression \cite{tin} 
\begin{equation}
H_{c1}(0) = \frac{\Phi_{0}}{4\pi\lambda_{GL}^2(0)}\left(\mathrm{ln}\frac{\lambda_{GL}(0)}{\xi_{GL}(0)}\right).  
\label{eqn3:ld}
\end{equation}
Integrating the values of $H_{c1}$(0) = (2.98 $\pm$ 0.11) mT and $\xi_{GL}$(0) = (99 $\pm$ 1) \text{\AA} in Eq. \eqref{eqn3:ld}, yields $\lambda_{GL}$(0) = (4608 $\pm$ 88) $\text{\AA}$. The Ginzburg-Landau parameter $\kappa$ = $\lambda_{GL}$/$\xi_{GL}$ = 46.5 $\pm$ 0.4 $>$ 1/$\sqrt{2}$, therefore, NbOs$_{2}$ classified as strong type-II superconductor.\\
The thermodynamic critical field $H_{c}$(0) is given by 
\begin{equation}
H_{c}(0) = \frac{\Phi_{0}}{2\sqrt{2}\pi\lambda_{GL}\xi_{GL}(0)},  
\label{eqn4:tcf}
\end{equation}
and found to be $H_{c}$(0) = (51 $\pm$ 1) mT.\\
Recently unconventional vortex dynamics has been observed in some noncentrosymmetric superconductors where unusual vortex pinning mechanism  and flux creep rates were proposed \cite{CFM1,CFM2}. Therefore, it is necessary to measure the stability of vortex system against the various factors unsettling the vortex equilibrium. The measure of stability against the thermal fluctuations can be given by Ginzburg number $G_{i}$. This is basically the ratio of thermal energy $k_{B}$$T_{c}$ to the condensation energy associated with the coherence volume. Ginzburg number, $G_{i}$, is given by \cite{GB}
\begin{equation}
 G_{i} = \frac{1}{2}\left(\frac{k_{B}\mu_{0}\tau T_{c}}{4\pi\xi^{3}(0)H_{c}^{2}(0)}\right)^2 ,
\label{eqn7:gi}
\end{equation}
where $\tau$ is anisotropy parameter which is 1 for the cubic NbOs$_{2}$. For $\xi$(0) = (99 $\pm$ 1) \text{\AA}, $H_{c}$(0) = (51 $\pm$ 1) mT and $T_{c}$ = (2.7 $\pm$ 0.1)  K, we obtained $G_{i}$ = (1.1 $\pm$ 0.1) $\times$ 10$^{-6}$. In conventional low $T_{c}$ superconductors, pinning is strong, whereas thermal fluctuations are weak, with $G_{i}$ $\sim$ 10$^{-8}$. In high temperature superconductors, $T_{c}$ is high and hence the coherence volume is small, making it sensitive to thermal fluctuations, $G_{i}$ $\sim$ 10$^{-2}$. In our compound $G_{i}$ value is more towards the low $T_{c}$ superconductors, suggesting that thermal fluctuations may not be playing any important role in vortex unpinning in this material.\\ 
Figure \ref{Fig4:cp} shows the low temperature specific heat data measured in zero applied field. The plot $C/T$ vs $T^{2}$ is shown in the inset of \figref{Fig4:cp}, in the temperature range 3 K $\le$ $T^{2}$ $\le$ 100 K. A jump at around $T_{c}$ = 2.7 K, apparently confirms bulk superconductivity in NbOs$_{2}$. The normal state specific heat data above $T_{c}$ is contributed by both electronic and phononic parts given by: $C/T = \gamma_{n}+\beta_{3}T^{2}+\beta_{5}T^{4}$, where $\gamma_{n}$ is the Sommerfeld coefficient and $\beta_{3}$, $\beta_{5}$ are  phononic contributions. The solid red line represents the best fit to the data with $\gamma_{n}$ = 8.58 $\pm$ 0.01 mJ mol$^{-1}$ K$^{-2}$, $\beta_{3}$ = 0.123 $\pm$ 0.002 mJ mol$^{-1}$ K$^{-4}$, and $\beta_{5}$ = 0.43 $\pm$ 0.04 $\mu$J mol$^{-1}$ K$^{-6}$ [see inset \figref{Fig4:cp}]. The value for $\gamma_{n}$ was used to determine the density of states at the Fermi level $D_{c}(E_{\mathrm{F}})$ using the relation $\gamma_{n} =( \pi^{2}k_{B}^{2}D_{c}(E_{\mathrm{F}}))/3$, where $E_{\mathrm{F}}$ is the Fermi energy.  For  $\gamma_{n}$ = 8.58 $\pm$ 0.01 mJ mol$^{-1}$ K$^{-2}$, it yields $D_{c}(E_{\mathrm{F}})$ = 3.64 $\pm$ 0.02 states eV$^{-1}$ f.u.$^{-1}$. The Debye temperature $\theta_{D}$ = $\left(12\pi^{4}RN/5\beta_{3}\right)^{1/3}$, where using R = 8.314 J mol$^{-1}$ K$^{-1}$ and N = 3, yields $\theta_{D}$ = 362 $\pm$ 1 K. The value of $\theta_{D}$ = 362 K can be used to calculate the electron-phonon coupling constant $\lambda_{e-ph}$  using the McMillan formula \cite{WL},
\begin{equation}
\lambda_{e-ph} = \frac{1.04+\mu^{*}\mathrm{ln}(\theta_{D}/1.45T_{c})}{(1-0.62\mu^{*})\mathrm{ln}(\theta_{D}/1.45T_{c})-1.04 } ,
\label{eqn4:ld}
\end{equation}                       
where $\mu^{*}$ represents the repulsive screened Coulomb potential, usually given by $\mu^{*}$ = 0.13. With $T_{c}$ = 2.7 K and $\theta_{D}$ = 361.67 K, we obtained $\lambda_{e-ph}$ $\simeq$ 0.52, which is similar to other weakly coupled NCS superconductors \cite{ABK,SAS}.\\ 
\begin{figure}
\includegraphics[width=1.0\columnwidth]{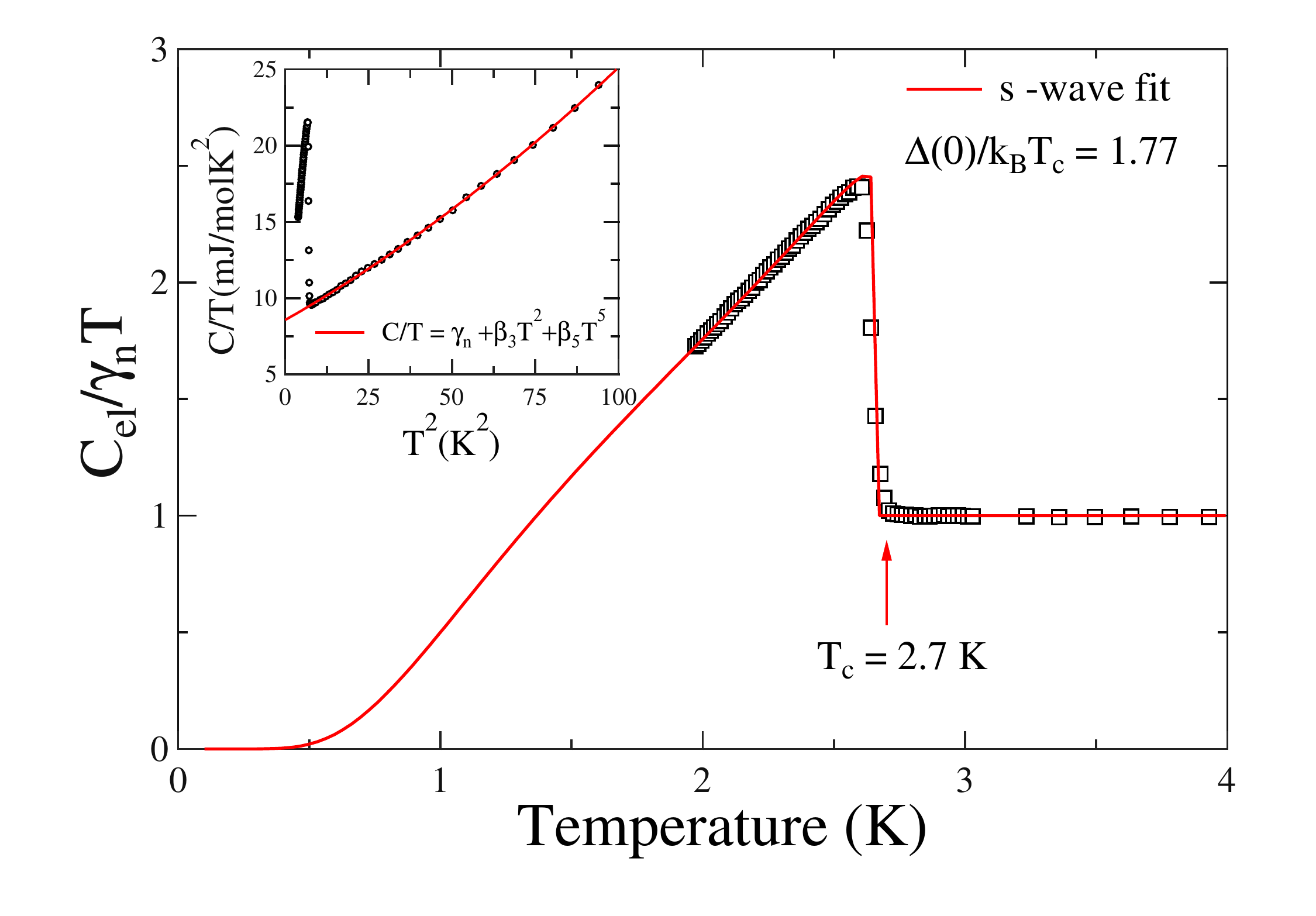}
\caption{\label{Fig4:cp} The specific heat data in superconducting regime fitted for single-gap s-wave model [Eq.\eqref{eqn6:Cel}] for a fitting parameter $\alpha$ = $\Delta(0)/k_{B}T_{c}$ = 1.77 $\pm$ 0.01. Inset: $C/T$ vs $T^{2}$ data was fitted between 3 K $\le$ $T^{2}$ $\le$ 100 K by Eq. $C/T = \gamma_{n}+\beta_{3}T^{2}+\beta_{5}T^{4}$ to measure electronic and phononic contribution to specific heat.}
\end{figure} 
The electronic contribution ($C_{el}$) to the specific heat  can be calculated by subtracting the phononic contribution ($C_{ph}$) from total specific heat data.
\begin{equation}
C_{el} = C-C_{ph} = C-(\beta_{3}T^{3}+\beta_{5}T^{5}).
\label{eqn5:cel}
\end{equation}
Once the phononic contribution is subtracted, an equal entropy conservation line is drawn to estimate the normalized specific heat jump. The value for the specific heat jump $\frac{\Delta C}{\gamma_{n}T_{c}}$  $\approx$ 1.45, which is remarkebly  similar to the value for a BCS isotropic gap superconductor (= 1.43) in the weak coupling limit, indicating that NbOs$_{2}$ is a weakly coupled superconductor, consistent with the $\lambda_{e-ph}$ value obtained above.\\
\begin{figure}
\includegraphics[width=1.0\columnwidth]{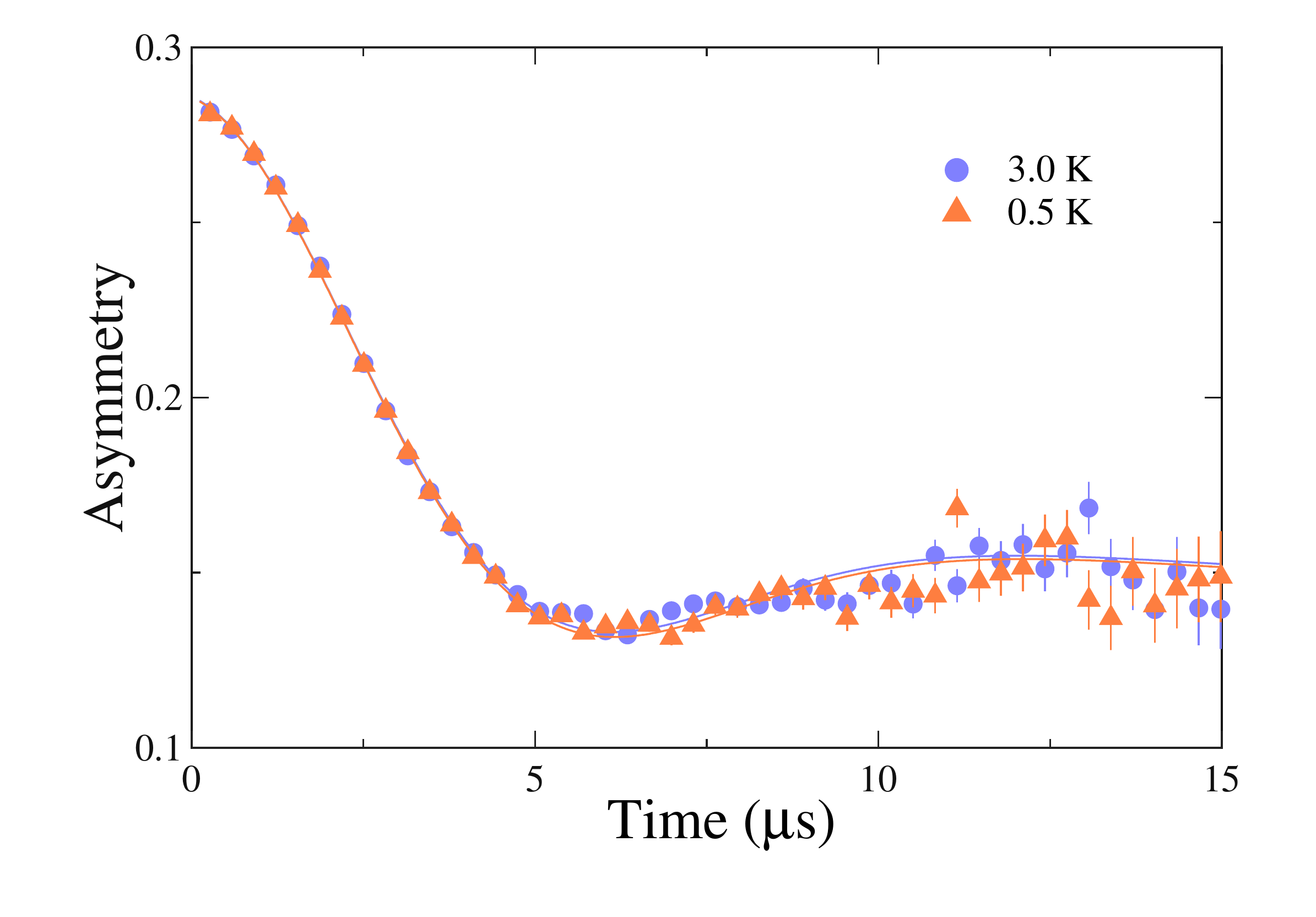}
\caption{\label{Fig5:ZFM} Zero-field $\mu$SR spectra collected below (0.5 K) and above (3 K) the superconducting transition temperature. The solid lines are the fits to Guassian Kubo-Toyabe (KT) function given in Eq.\eqref{eqn8:zf}.}
\end{figure}
The temperature dependence of the specific heat data in the superconducting state can best be described by the BCS expression for the normalized entropy S written as
\begin{equation}
\frac{S}{\gamma_{n}T_{c}} = -\frac{6}{\pi^2}\left(\frac{\Delta(0)}{k_{B}T_{c}}\right)\int_{0}^{\infty}[ \textit{f}\ln(f)+(1-f)\ln(1-f)]dy ,
\label{eqn5:s}
\end{equation}
where $\textit{f}$($\xi$) = [exp($\textit{E}$($\xi$)/$k_{B}T$)+1]$^{-1}$ is the Fermi function, $\textit{E}$($\xi$) = $\sqrt{\xi^{2}+\Delta^{2}(t)}$, where $\xi$ is the energy of normal electrons measured relative to the Fermi energy, $\textit{y}$ = $\xi/\Delta(0)$, $\mathit{t = T/T_{c}}$, and $\Delta(t)$ = tanh[1.82(1.018(($\mathit{1/t}$)-1))$^{0.51}$] is the BCS approximation for the temperature dependence of the energy gap. The normalized electronic specific heat is calculated by
\begin{equation}
\frac{C_{el}}{\gamma_{n}T_{c}} = t\frac{d(S/\gamma_{n}T_{c})}{dt}.
\label{eqn6:Cel}
\end{equation}
Fitting the low temperature specific heat data using this model as shown by the solid red line in \figref{Fig4:cp}, yields $\alpha$ = $\Delta(0)/k_{B}T_{c}$ = 1.77 $\pm$ 0.01. This is consistent with the value for a BCS superconductor $\alpha_{BCS}$ = 1.764 in the weak coupling limit. Therefore, good agreement between the measured data (black symbols) and the BCS fit (solid red line), confirms an isotropic fully gapped BCS superconductivity in NbOs$_{2}$. It should be noted that in order to extract the true nature of  the superconducting gap, the specific heat data up to the low temperature region need to be analyzed. It can give a complete understanding whether the superconducting gap is isotropic (exponential) or have nodes (power law). A summary of all the experimentally measured and estimated parameters is given in Table \ref{superconducting properies}. \\
\begin{figure}[t]
\includegraphics[width=1.0\columnwidth]{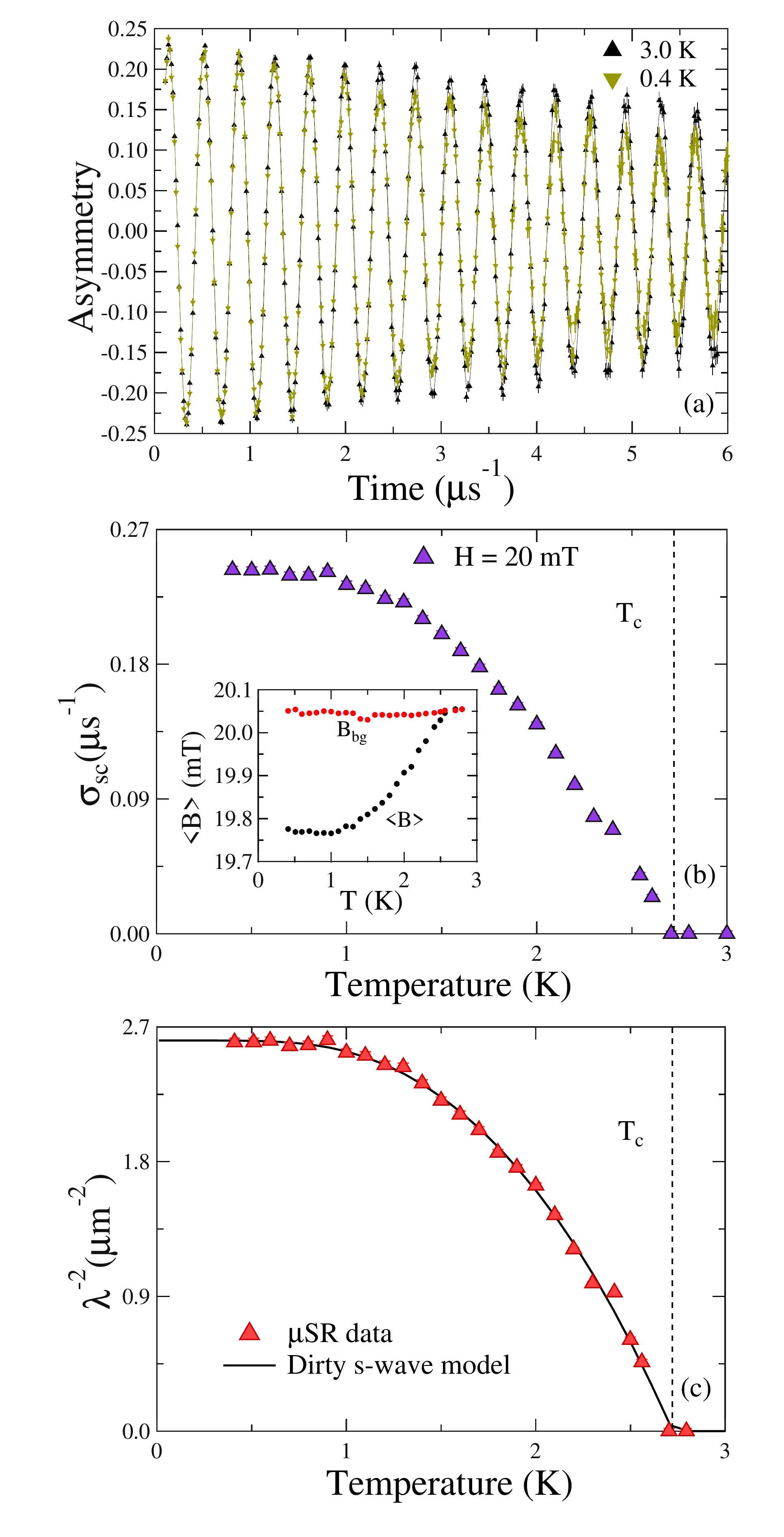}
\caption{\label{Fig6:TF}(a) Representative TF-$\mu$SR signals collected at 3.0 K and 0.4 K in an applied magnetic field of 20 mT.(b) Temperature dependence of the superconducting contribution to depolarization $\sigma_{\mathrm{sc}}$ in an applied magnetic field of 20 mT. The inset shows the internal field variation with temperature where a clear diamagnetic signal appears around $T_{c}$. (c) Temperature dependence of $\lambda^{-2}$ follow a single gap s-wave model in dirty limit for $\Delta(0)$ = 0.43 $\pm$ 0.02.}
\end{figure}
To further inspect the superconducting ground state of NbOs$_{2}$, we performed the muon spin relaxation and rotation measurements. Firstly, ZF-$\mu$SR measurements were carried out to investigate the occurrence of TRS breaking in NbOs$_{2}$. ZF-$\mu$SR being extremely sensitive to tiny changes in the internal magnetic fields, can unambiguously detect the presence of a TRS breaking signal. This technique was proved to be very crucial in establishing the TRS breaking in several superconductors such as La$_{7}$Ir$_{3}$ \cite{JAT}, Re$_{6}$(Ti,Hf,Zr) \cite{RPS,DSJ,DSJ1},  Sr$_{2}$RuO$_{4}$ \cite{SRO1,SRO2} etc. Figure \ref{Fig5:ZFM} shows the muon spin relaxation time spectra, collected below ($T = 0.5$ K < $T_{c}$) and above ($T = 3.0$ K > $T_{c}$) the superconducting transition temperature $T_{c}$ = 2.7 K. The absence of precessional signals suggests the absence of coherent internal fields which is generally associated with long range magnetic ordering. In the absence of atomic moments muon-spin relaxation in zero field can best be described by the Gaussian Kubo-Toyabe (KT) function \cite{RSH} together with  a non-decaying constant background,  $A_{\mathrm{BG}}$, 
\begin{eqnarray}
G_{\mathrm{KT}}(t) &=& A_{1}\left[\frac{1}{3}+\frac{2}{3}(1-\sigma^{2}_{\mathrm{ZF}}t^{2})\mathrm{exp}\left(\frac{-\sigma^{2}_{\mathrm{ZF}}t^{2}}{2}\right)\right]\mathrm{exp(-\Lambda t)}\nonumber\\&+&A_{\mathrm{BG}} ,
\label{eqn8:zf}
\end{eqnarray} 
where $A_{1}$ is the initial asymmetry and $\sigma_{\mathrm{ZF}}$ accounts for the relaxation due to static, randomly oriented local fields associated with the nuclear moments at the muon site and $\Lambda$ is the electronic relaxation rate. Near identical relaxation signals,  as seen in Figure \ref{Fig5:ZFM}, on the either side of the superconducting transition suggest the absence of any additional magnetic moments in the superconducting ground state, usually associated with exotic phenomena such as TRS breaking. This clearly suggests that the time-reversal symmetry is preserved in NbOs$_{2}$ within the detection limit of $\mu$SR.\\
Figure \ref{Fig6:TF}(a) shows the TF-$\mu$SR precessional signals for NbOs$_{2}$ collected above (3.5 K) and below (0.4 K) $T_{c}$ in an applied magnetic field of 20 mT. The measurements were done in the field-cooled mode where the field 20 mT was applied above the transition temperature. The sample was then subsequently cooled to the base temperature of 0.4 K. 
In the normal state there is an almost homogeneous field distribution throughout the sample, where the weak depolarization is due to nuclear dipolar field. The depolarization rate in the superconducting state becomes more prominent, due to the formation of an inhomogeneous field distribution in the flux line lattice (FLL) state. The TF-$\mu$SR spectra in  \ref{Fig6:TF}(a) show a very small difference between above and below $T_{c}$ , indicating large magnetic penetration depth.\\
The TF-$\mu$SR time spectra, for all temperatures, is best described by the sinusoidal oscillatory function damped with a Gaussian relaxation and an oscillatory background term arising from the muons implanted directly into the silver sample holder that do not depolarize,
\begin{eqnarray}
G_{\mathrm{TF}}(t) &=& A_{0}\mathrm{exp}\left(\frac{-\sigma^{2}t^{2}}{2}\right)\mathrm{cos}(\gamma_{\mu}B_{int}t+\phi)\nonumber\\&+&A_{1}\mathrm{cos}(\gamma_{\mu}B_{bg}t+\phi) .
\label{eqn14:Tranf}
\end{eqnarray}
Here $A_{0}$ and $A_{1}$ are the initial asymmetries of the sample and background signals, $B_{int}$ and $B_{bg}$ are the internal and background magnetic fields, $\phi$ is the initial phase offset, $\gamma_{\mu}/2\pi$ = 135.53 MHz/T is the  muon gyromagnetic ratio and $\sigma$ is the depolarization rate. The depolarization rate $\sigma$ comprised of the following terms: $\sigma^{2}$ = $\sigma_{\mathrm{sc}}^{2}+\sigma_{\mathrm{N}}^{2}$, where $\sigma_{\mathrm{sc}}$ is the depolarization arising due to the field variation across the flux line lattice and $\sigma_{\mathrm{N}}$ is the contribution due to nuclear dipolar moments. The superconducting contribution to the depolarization rate $\sigma_{\mathrm{sc}}$ is calculated by using the above relation. Figure \ref{Fig6:TF}(b) shows the temperature dependence of $\sigma_{\mathrm{sc}}$ where below $T_{c}$, $\sigma_{\mathrm{sc}}$ increases systematically as the temperature is lowered. The inset of \figref{Fig6:TF}(b) shows the temperature dependence of internal magnetic field,  where the flux expulsion at $T_{c}$ is evident from the reduction of average feld $<B>$ inside the superconductor, and the corresponding background field $B_{bg}$ is approximately constant over the temperature range. 
In an isotropic type-II superconductor with a hexagonal Abrikosov vortex lattice the magnetic penetration depth $\lambda$ is related to $\sigma_{\mathrm{sc}}$ by the equation \cite{EHB}:
\begin{equation}
\sigma_{\mathrm{sc}}(\mu s^{-1}) = 4.854 \times 10^{4}(1-h)[1+1.21(1-\sqrt{h})^{3}]\lambda^{-2}, 
\label{eqn3:sigmaH}
\end{equation}
where $h = H/H_{c2}(T)$ is the reduced field. The above equation is valid for  systems $\kappa > 5$. For NbOs$_{2}$, $\kappa$ = (46.5 $\pm$ 0.4) and the temperature dependence of $H_{c2}(T)$ is shown in \figref{Fig3:hc1}(d). Using the data of $H_{c2}(T$) for NbOs$_{2}$, the temperature dependence of $\lambda^{-2}$ was extracted from \equref{eqn3:sigmaH}, as displayed in \figref{Fig6:TF}(c). The temperature dependence of $\lambda^{-2}$ nearly constant below $T_{c}$/3 $\approx$ 0.9 K. This possibly suggests the absence of nodes in the superconducting energy gap at the Fermi surface. The solid line in \figref{Fig6:TF}(c) represent the temperature dependence of the London magnetic penetration depth $\lambda(T)$ within the local London approximation for a s-wave BCS superconductor in the dirty limit:
\begin{equation}
\frac{\lambda^{-2}(T)}{\lambda^{-2}(0)} = \frac{\Delta(T)}{\Delta(0)}\mathrm{tanh}\left[\frac{\Delta(T)}{2k_{B}T}\right] ,
\label{eqn4:lpd}
\end{equation}
where $\Delta(T)/\Delta(0) = \tanh\{1.82(1.018({T_{c}/T}-1))^{0.51}\}$ is the BCS approximation for the temperature dependence of the energy gap, where $\Delta(0)$ is the gap magnitude at zero temperature. Dirty-limit expression was used in accordance with the calculation done in the section below where we found $\xi_{0}$ > $\textit{l}$. The fit yields energy gap $\Delta (0)$ = (0.43 $\pm$ 0.02) meV and $\lambda$ (0) = (6190 $\pm$ 45) \text{\AA}. The gap to $T_{c}$ ratio $\Delta (0)/k_{B}T_{c}$ = 1.82 is close to the value of 1.764 expected from the BCS theory, suggesting that NbOs$_{2}$ is a weakly-coupled superconductor. 
\begin{table}[h!]
\caption{Experimentally measured and estimated superconducting and normal-state properties for the noncentrosymmetric superconductor NbOs$_{2}$}
\label{superconducting properies}
\begin{center}
\begin{tabular*}{1.0\columnwidth}{l@{\extracolsep{\fill}}lll}\hline\hline
Properties& unit& value\\
\hline
\\[0.5ex]                                  
$T_{c}$& K& 2.7 $\pm$ 0.2\\             
$H_{c1}(0)$& mT& 2.98 $\pm$ 0.11 \\                       
$H_{c2}(0)$& T& 3.36 $\pm$ 0.07 \\
$H_{c}(0)$& mT& 51 $\pm$ 1 \\
$H_{c2}^{P}(0)$& T& 5.02 $\pm$ 0.18\\
$\xi_{GL}$& \text{\AA}& 99 $\pm$ 1\\
$\lambda_{GL}$& \text{\AA}& 4608 $\pm$ 88\\
$\kappa$& &46.5 $\pm$ 0.4\\
$\gamma$& mJmol$^{-1}$K$^{-2}$& 8.58 $\pm$ 0.01\\
$\beta$ & mJmol$^{-1}$K$^{-4}$& 0.123 $\pm$ 0.002\\
$\theta_{D}$& K& 362 $\pm$ 1\\
$\lambda_{e-ph}$&  &0.52 $\pm$ 0.02\\
D$_{C}$(E$_{f}$)& states/ev f.u& 3.64 $\pm$ 0.02\\
$\Delta C_{el}/\gamma_{n}T_{c}$&   &1.45 $\pm$ 0.03\\
$\Delta(0)/k_{B}T_{c}$&   &1.77 $\pm$ 0.01
\\[0.5ex]
\hline\hline
\end{tabular*}
\par\medskip\footnotesize
\end{center}
\end{table} 
\section{Electronic properties}
The Sommerfeld coefficient is related to the quasiparticle number density (n) per unit volume given by the relation \cite{ck}
\begin{equation}
\gamma_{n} = \left(\frac{\pi}{3}\right)^{2/3}\frac{k_{B}^{2}m^{*}V_{\mathrm{f.u.}}n^{1/3}}{\hbar^{2}N_{A}}
\label{eqn14:gf}
\end{equation}
where k$_{B}$ is the Boltzmann constant, N$_{A}$ is the Avogadro number, V$_{\mathrm{f.u.}}$ is the volume of a formula unit and m$^{*}$ is the effective mass of quasiparticles. The residual resistivity, $\rho_{0}$, can be calculated, using the equation
\begin{equation}
\textit{l} = \frac{3\pi^{2}{\hbar}^{3}}{e^{2}\rho_{0}m^{*2}v_{\mathrm{F}}^{2}}
\label{eqn15:le}
\end{equation}
while the Fermi velocity $v_{\mathrm{F}}$ can be written in terms of  effective mass and the carrier density by
\begin{equation}
n = \frac{1}{3\pi^{2}}\left(\frac{m^{*}v_{\mathrm{F}}}{\hbar}\right)^{3} .
\label{eqn16:n}
\end{equation} 
For superconductors in the dirty-limit, where $\xi_{0}/l$ $>>$ 1, the properties are affected due to the scattering of electrons. The dirty limit expression for the penetration depth $\lambda$(0) is then given by \cite{tin}
\begin{equation}
\lambda(0) = \lambda_{L}\left(1+\frac{\xi_{0}}{\textit{l}}\right)^{1/2}
\label{eqn17:f}
\end{equation}
where $\xi_{0}$ is the BCS coherence length. The  London penetration depth, $\lambda_{L}$, is given by
\begin{equation}
\lambda_{L} = \left(\frac{m^{*}}{\mu_{0}n e^{2}}\right)^{1/2}
\label{eqn18:laml}
\end{equation}
The BCS coherence length $\xi_{0}$ and the Ginzburg-Landau coherence, $\xi_{GL}$(0), at T = 0 K in the dirty limit is related by the expression
\begin{equation}
\frac{\xi_{GL}(0)}{\xi_{0}}= \frac{\pi}{2\sqrt{3}}\left(1+\frac{\xi_{0}}{l}\right)^{-1/2}
\label{eqn19:xil}
\end{equation}
Using Eq.\eqref{eqn14:gf}-\eqref{eqn19:xil} form a system of equations which can be used to estimate the parameters m$^{*}$, n, $\textit{l}$, and $\xi_{0}$ as done in Ref.\cite{DAM,JAT1}. The system of equations was solved simultaneously using the values $\gamma_{n}$ = (8.58 $\pm$ 0.01) mJ mol$^{-1}$K$^{-2}$, $\xi(0)$ = (99 $\pm$ 1) \text{\AA} and $\rho_{0} $ = (130 $\pm$ 5) $ \mu\Omega $cm. The values were estimated for $\lambda^{H_{c1}}$ $\simeq$ 4608 \text{\AA} and $\lambda^{\mu}$ $\simeq$ 6190 \text{\AA}, tabulated in Table \ref{elec propr}. It is clear that $\xi_{0}$ > $\textit{l}$, indicating that NbOs$_{2}$ is in the dirty limit as previously asserted. This also accounts for the low RRR and residual resistivity that have been measured.
\begin{figure}[t]
\includegraphics[width=1.0\columnwidth]{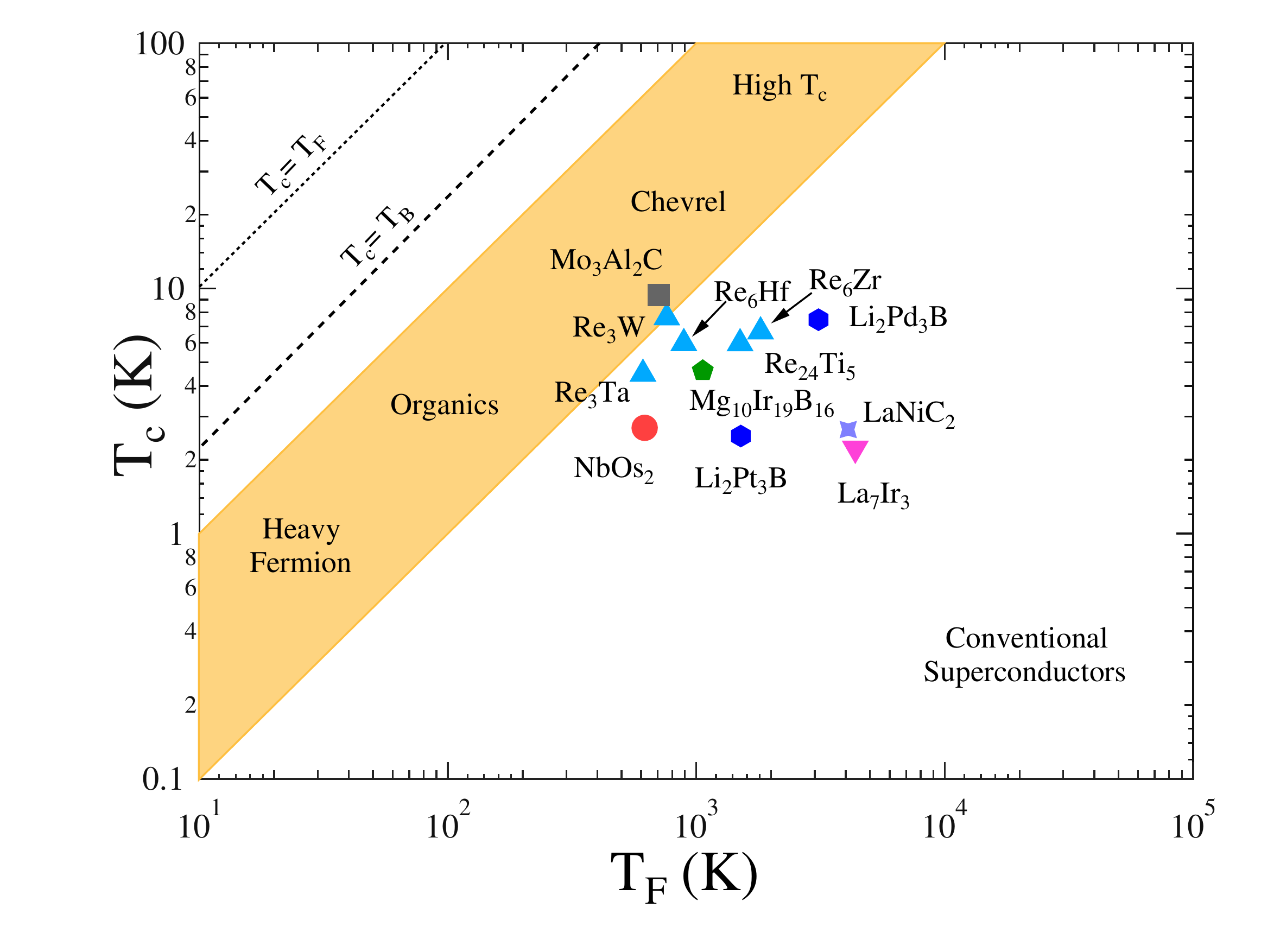}
\caption{\label{Fig7:up} The Uemura plot showing the superconducting transition temperature $T_{c}$ vs the Fermi temperature $T_{F}$, where NbOs$_{2}$ is shown as a solid red circle just outside the range of band of unconventional superconductors.  $T_{B}$ is the Bose-Einstein condensation temperature.  The orange region represents the band of unconventionality, other points with broken TRS such as LaNiC$_{2}$, La$_{7}$Ir$_{3}$ etc. and Re-based $\alpha$-$Mn$ structure compounds   were obtained from Ref. \cite{JAT1,YJU1,YJU2,YJU3}.}
\end{figure}
The bare-band effective mass m$^{*}_{band}$ can be related to m$^{*}$, which contains enhancements from the many-body electron phonon interactions \cite{GG} m$^{*}$ = m$^{*}_{band}$(1+$\lambda_{e-ph}$), where $\lambda_{e-ph}$ is the electron-phonon coupling constant. Using this value of $\lambda_{e-ph}$ = 0.52, a value for m$^{*}_{band}$ can be found, given in Table \ref{elec propr}.\\ 
\begin{table}[h!]
\caption{Electronic properties calculated of NbOs$_{2}$ for $\lambda^{H_{c1}}$ $\simeq$ 4608 \text{\AA} and $\lambda^{\mu}$ $\simeq$ 6190 \text{\AA}}
\label{elec propr}
\begin{center}
\begin{tabular*}{1.0\columnwidth}{l@{\extracolsep{\fill}}lll}\hline\hline
Properties& unit& $H_{c1}$& $\mu SR$\\
\hline
\\[0.5ex]                                  
$\lambda_{GL}$& \text{\AA}& 4608& 6190 \\
$m^{*}/m_{e}$&  &11.6 $\pm$ 0.3 &13.4$\pm$ 0.3\\
$m^{*}_{band}/m_{e}$&  &7.63 $\pm$ 0.13&8.8 $\pm$ 0.2\\
$n$& 10$^{27}$m$^{-3}$& 4.16 $\pm$ 0.23 & 2.74 $\pm$ 0.17\\
$\xi_{0}$& \text{\AA}& 64 $\pm$ 5 & 90 $\pm$ 6\\
$l$& \text{\AA}& 37$\pm$ 4& 50 $\pm$ 3\\
$\lambda_{L}$& \text{\AA} & 2809 $\pm$ 130& 3720 $\pm$ 164 \\
$v_{\mathrm{F}}$& 10$^{4}$m s$^{-1}$& 4.95 $\pm$ 0.23& 3.75 $\pm$ 0.16\\
$T_{\mathrm{F}}$& K& 1060 $\pm$ 30& 620 $\pm$ 20\\
$T_{c}/T_{\mathrm{F}}$& & 0.0025 $\pm$ 0.0001& 0.0044 $\pm$ 0.0001\\
\\[0.5ex]
\hline\hline
\end{tabular*}
\par\medskip\footnotesize
\end{center}
\end{table} 
\section{Uemura Plot}
For a 3D system the Fermi temperature T$_{F}$ is given by the relation
\begin{equation}
 k_{B}T_{F} = \frac{\hbar^{2}}{2}(3\pi^{2})^{2/3}\frac{n^{2/3}}{m^{*}}, 
\label{eqn13:tf}
\end{equation}
where n is the quasiparticle number density per unit volume.
According to Uemura et al. \cite{YJU1,YJU2,YJU3} superconductors can be conveniently classified according to their $\frac{T_{c}}{T_{F}}$ ratio. It was shown that for the unconventional superconductors such as heavy-fermion, high- $T_{c}$, organic superconductors, and iron-based superconductors this ratio falls in the range 0.01 $\leq$ $\frac{T_{c}}{T_{F}}$ $\leq$ 0.1. In \figref{Fig7:up}, the orange region represents the band of unconventional superconductors. Uemura plot is presented in accordance with Ref. \cite{JAT1,YJU1,YJU2,YJU3}, where it shows the superconductors with unconventional properties such as Mo$_{3}$Al$_{2}$C, Mg$_{10}$Ir$_{19}$B$_{16}$ Li$_{2}$Pt$_{3}$B, LaNiC$_{2}$, La$_{7}$Ir$_{3}$ and the family of Re-based $\alpha$-$Mn$ structure superconductors with broken TRS. Using the estimated value of n and m$^{*}$ for NbOs$_{2}$ in Eq. \ref{eqn13:tf}, it yields $T_{F}$ = 620 K, giving $\frac{T_{c}}{T_{F}}$ = 0.004. NbOs$_{2}$ is located just outside the range of unconventional superconductors as shown by a solid red marker in \figref{Fig7:up}, potentially borders an unconvnetional superconducting ground state.  Interestingly, all the Re$_{6}$X  \cite{RPS,DSJ,DSJ1} noncentrosymmetric superconductors with TRS breaking are located in the same phase space in the Uemura plot. The close proximity of the entire family might also be pointing towards the common origin of TRS breaking, which seems to be Re SOC. However,  other Re-based compounds for example Re$_{3}$X (X =W, Ta)\cite{PKB,JAT1} show preserved TRS. This may be due to the reduced Re composition in these compounds which may have a major effect on the underlying electronic characteristics of the spin-triplet channel. In addition, our ZF-$\mu$SR measurements on the Re-free NbOs$_{2}$ also advocate the above conclusion, since TRS is found to be preserved in the superconducting state of this material. Therefore, this makes the role of Re element even more interesting. However, to make a generic comment about the contribution of Re in the observed TRS breaking in this family, it is important to study more Re-rich $\alpha$-$Mn$ materials and more importantly Re itself.

\section{Summary and Conclusions}
Detailed investigation of the normal and superconducting phase properties of NbOs$_{2}$ was carried out using XRD, magnetization, resistivity, specific heat and $\mu$SR measurements. Sample was prepared by standard arc-melting technique, where the phase purity and noncentrosymmetric $\alpha$-$Mn$ structure for NbOs$_{2}$ was confirmed by XRD analysis. Our results suggest type-II dirty limit superconductivity in NbOs$_{2}$ with superconducting transition temperature $T_{c}^{onset}$ = 2.7 $\pm$ 0.1 K. The specific heat measurements confirm the superconducting gap is isotropic and are, within error, the same as the BCS predicted values. ZF-$\mu$SR measurement confirmed that time-reversal symmetry is preserved, within the detection limit of $\mu$SR. The TF-$\mu$SR measurements also confirm fully gapped BCS superconductivity with no point or line nodes. In addition, in the Uemura plot NbOs$_{2}$ is placed just outside the borderline of the band of unconventionality. Re-based $\alpha$-$Mn$ structure superconductors with broken TRS are situated very close to each other in the Uemura plot. The negligible effect of ASOC and persistence occurrence of TRS breaking in this family of compounds hints at its common origin or mechanism. It could be due to the Re SOC which facilitates the origin of TRS breaking in this family. The above conclusion is supported by the absence of TRS breaking phenomena in Re-free NbOs$_{2}$ having the same structure. Further confirmation is provided by the recent observation of TRS breaking in pure Re metal (centrosymmetric). Therefore, it is imperative and timely to search for TRS breaking in other Re-rich superconductors and employ both experimental and theoretical approach to identify the origin of TRS breaking.  

\section{Acknowledgments}
R.~P.~S.\ acknowledges Science and Engineering Research Board, Government of India for the Young Scientist Grant No. YSS/2015/001799 and Department of Science and Technology, India (SR/NM/Z-07/2015) for the financial support and Jawaharlal Nehru Centre for Advanced Scientific Research (JNCASR) for managing the project. We thank ISIS, STFC, UK for the beamtime to conduct the $\mu$SR experiments

\end{document}